\documentclass[pra,amsmath,amssymb,showpacs,superscriptaddress,twocolumn]{revtex4}

\usepackage{graphicx}
\usepackage{dcolumn}
\usepackage[hypertex]{hyperref}
\usepackage{bm}
\usepackage[usenames,dvipsnames]{color}
\usepackage{amsmath}
\usepackage{amsfonts}
\usepackage{amsthm}
\usepackage{amssymb}
\usepackage{mathrsfs}
\usepackage{subfigure}
\usepackage{ulem}

\begin{document}

\title{Multiple phase estimation in quantum cloning machines}

\author{Yao Yao}
\affiliation{Beijing Computational Science Research Center, Beijing, 100084, China}

\affiliation{Synergetic Innovation Center of Quantum Information and Quantum Physics, University of Science and Technology of China,
Hefei, Anhui 230026, China}

\author{Li Ge}
\affiliation{Beijing Computational Science Research Center, Beijing, 100084, China}

\author{Xing Xiao}
\affiliation{Beijing Computational Science Research Center, Beijing, 100084, China}

\author{Xiao-guang Wang}
\email{xgwang@zimp.zju.edu.cn}
\affiliation{Zhejiang Institute of Modern Physics, Department of Physics, Zhejiang University, Hangzhou 310027, China}

\author{Chang-pu Sun}
\email{cpsun@csrc.ac.cn}
\affiliation{Beijing Computational Science Research Center, Beijing, 100084, China}
\affiliation{Synergetic Innovation Center of Quantum Information and Quantum Physics, University of Science and Technology of China,
Hefei, Anhui 230026, China}

\date{\today}

\begin{abstract}
Since the initial discovery of the Wootters-Zurek no-cloning theorem, a wide variety of quantum cloning machines have been proposed aiming at
imperfect but \textit{optimal} cloning of quantum states within its own context. Remarkably, most previous studies have employed the Bures fidelity or the Hilbert-Schmidt
norm as the figure of merit to characterize the quality of the corresponding cloning scenarios. However, in many situations, what we truly
care about is the relevant information about certain parameters encoded in quantum states. In this work, we investigate the multiple phase estimation
problem in the framework of quantum cloning machines, from the perspective of quantum Fisher information matrix (QFIM). Focusing on the generalized
$d$-dimensional equatorial states, we obtain the analytical formulas of QFIM for both universal quantum cloning machine (UQCM) and phase-covariant
quantum cloning machine (PQCM), and prove that PQCM indeed performs better than UQCM in terms of QFIM. We highlight that our method can be generalized
to arbitrary cloning schemes where the fidelity between the single-copy input and output states is input-state independent. Furthermore,
the attainability of the quantum Cram\'{e}r-Rao bound is also explicitly discussed.

\end{abstract}


\maketitle
\section{INTRODUCTION}
The no-cloning theorem, initially discovered in the early 1980s, is one of the earliest and paramount results of quantum computation and quantum information,
which prohibits the probability of perfectly cloning an arbitrary unknown state \cite{Wootters1982,Dieks1982}. However, approximate
or probabilistic cloning can still be accomplished with new conceptual and technical tools developed within the framework of quantum information theory \cite{Scarani2005}.
Since then, many refinements of the no-cloning theorem and various quantum cloning machines have been proposed, such as Wootters-Zurek cloning \cite{Wootters1982},
universal cloning \cite{Buzek1996,Gisin1997,Buzek1998}, state-dependent cloning \cite{Bruss1998}, probabilistic cloning \cite{Duan1998a,Duan1998b},
phase-covariant cloning \cite{Niu1999,Bruss2000,DAriano2001,Fan2001,Cerf2002a,Fan2003,Buscemi2005,Durt2005}, just to name a few. All these schemes are \textit{optimal} in its own context,
where indicates some measures of distance metric are used to quantify the \textit{closeness} between the output copy and the input state. For instance,
the possible choices are the Uhlmann fidelity, the Bures distance, the Hilbert-Schmidt norm and the trace norm \cite{Kwek2000}. Moreover, it is worth emphasizing
that quantum cloning machines also find wide applications in other quantum information tasks \cite{Fan2013}.

On the other hand, in plenty of theoretical and experimental scenarios, our real concern is only the partial information about some relevant parameters encoded in quantum states
instead of the states themselves, as pointed out by Lu and Song \cite{Lu2013,Song2013}. Therefore, in these situations, all we need is to clone the relevant parameter information.
In order to quantify the physical information about these involved parameters, quantum Fisher information (QFI) is introduced \cite{Braunstein1994,Braunstein1996,Petz1996,Petz2011}
and receives more and more attention due to its great significance in both quantum estimation theory and quantum-enhanced metrology \cite{Giovanetti2004,Giovanetti2006,Pairs2009,Giovanetti2011}.
Remarkably, Lu \textit{et al.} investigated the cloning and broadcasting of QFI in a general sense and proved that QFI cannot be cloned \cite{Lu2013}. Furthermore, Song \textit{et al.}
compared the Wootters-Zurek cloning and universal cloning from the perspective of QFI and showed that the former performs better than the latter in this context \cite{Song2013}.
These results shed new light on the nature of QFI and can deepen our understanding of the information transferring in quantum cloning machines.

However, we note that Lu and Song only considered the \textit{single}-parameter case and cannot be directly extended to the multiple parameter case since the quantum
Cram\'{e}r-Rao bound (QCRB) cannot be generally saturated in multi-parameter problem \cite{Helstrom1976,Holevo1982}. On the other side, when we consider the cloning of $d$-dimensional
quantum system (especially for $d>2$), the multiple parameters are naturally involved such as phase-covariant quantum cloning of qudits \cite{Fan2003,Buscemi2005,Durt2005}.
These considerations motivate us to investigate the distributing and transferring of QFI in quantum cloning machines for qudits and to compare their performances in this particular context.
Quite recently, we also notice that the quantum estimation problem of multiple parameters is attracting increasing attention in the literature
\cite{Yuen1973,Helstrom1974,Belavkin1976,Fujiwara2001,Matsumoto2002,Ballester2004a,Ballester2004b,Chiribella2006,Imai2007,Genoni2008,Hayashi2008,Young2009,Watanabe2010,Monras2010,Monras2011,Genoni2013,
Gill2013,Vaneph2013,Humphreys2013,Crowley2014,Vidrighin2014}. With the aid of these results, we investigate the multiple phase estimation problem in quantum cloning machines for
qudits where universal quantum cloning machine (UQCM) and phase-covariant quantum cloning machine (PQCM) are both evaluated. Special focus is placed on the generalized $d$-dimensional
equatorial states \cite{Macchiavello2003,Mista2005} since this form of pure states has played a crucial role in many quantum information protocols such as quantum key distribution \cite{Bennett1984,Cerf2002b},
remote state preparation \cite{Bennett2001} and phase-covariant quantum cloning \cite{Fan2003,Buscemi2005,Durt2005}. We prove that PQCM indeed outperforms UQCM in terms of cloning QFI.
Moreover, the attainability of the quantum Cram\'{e}r-Rao bound and the generalization of our method are also discussed explicitly.

This paper is organized as follows. In Sec. \ref{preliminaries}, we provide a brief review of technical preliminaries of QFIM and its recent progress on the analytical calculation.
In Sec. \ref{calculations}, we discuss in detail the multi-parameter estimation problem in both UQCM and PQCM and give the analytical expressions of the corresponding QFIMs.
Furthermore, we illustrate that our method can be applied to a general class of quantum cloning machines. In Sec. \ref{attainability}, the attainability of the quantum Cram\'{e}r-Rao bound
is explicitly discussed. Finally, Sec. \ref{conclusion} is devoted to the discussion and conclusion.

\section{Technical preliminaries of QFIM}\label{preliminaries}
In this section, we will give a brief summary of multi-parameter estimation theory and review the recent progress on the analytical calculation of QFIM.
Let us consider a family of quantum states $\rho({\bm{\theta}})$ in the $d$-dimensional Hilbert space, involving a series of parameters denoted by
a vector $\bm{\theta}=\{\theta_\mu\}$, $\mu=1,\ldots,p$. For the single-parameter case (that is, $p=1$), the QFI is defined as \cite{Braunstein1994,Braunstein1996,Pairs2009}
\begin{equation}
\mathcal{F}(\theta)=\textrm{Tr}(\rho_\theta L_\theta^2),
\end{equation}
where the Hermite operator $L_\theta$ is the so called symmetric logarithmic derivative (SLD) satisfying \cite{Helstrom1976}
\begin{equation}
\frac{\partial\rho_\theta}{\partial\theta}=\frac{\rho_\theta L_{\theta}+ L_{\theta} \rho_\theta}{2},
\end{equation}
The quantum estimation theory places a fundamental limit on the estimation precision of the parameter $\theta$, which is characterized by
the QCRB
\begin{equation}
\mbox{Var}(\theta)\geq\frac{1}{M\mathcal{F(\theta)}}.
\end{equation}
Here $\mbox{Var}(\theta)$ denotes the variance of any unbiased estimator, and $M$ is the number of measurements repeated. It is worth
stressing that in this case the QCRB can always be asymptotically achieved with the maximum likelihood approach \cite{Helstrom1976}.

Turning to the multi-parameter scenario, the QFI is substituted by QFIM. The element of QFIM $\mathcal{F}(\bm{\theta})=[\mathcal{F}_{\mu\nu}]$
is defined by
\begin{equation}
\mathcal{F}_{\mu\nu}=\textrm{Tr}\left[\rho(\bm{\theta})\frac{L_\mu L_\nu+L_\nu L_\mu}{2}\right],
\end{equation}
where $L_\mu$ and $L_\nu$ are SLDs with respect to $\theta_\mu$ and $\theta_\nu$ respectively.
Meanwhile, the QCRB changes into the matrix inequality \cite{Helstrom1976}
\begin{equation}
\textrm{Cov}(\bm{\theta})\geq\big[M\mathcal{F(\bm{\theta})}\big]^{-1},
\end{equation}
where $\textrm{Cov}(\bm{\theta})$ stands for the covariance matrix of the estimator ${\bm{\hat\theta}}$.
Note that in general this bound cannot be achieved. Therefore, much effort has been devoted to the discussion of
the attainability of the multivariate QCRB. For pure states $\rho(\bm{\theta})=|\psi_{\bm{\theta}}\rangle\langle\psi_{\bm{\theta}}|$,
Fujiwara and Matsumoto proved that if the condition $\bm{\mbox{Im}}[\langle\psi_{\bm{\theta}}|L_\mu L_\nu|\psi_{\bm{\theta}}\rangle]=0$ is
satisfied for all $\mu$ and $\nu$, the multi-parameter QCRB is achievable at $\bm{\theta}$ \cite{Fujiwara2001,Matsumoto2002}.
Matsumoto also presented a POVM measurement with $p+2$ elements that indeed achieves the bound \cite{Matsumoto2002}. For mixed states,
the situation is more complicated. However, recent research by Gu\c{t}\u{a} and Kahn indicates that the QCRB is asymptotically
attainable if and only if \cite{Guta2007,Kahn2009,Guta}
\begin{equation}
\textrm{Tr}\big(\rho(\bm{\theta})[L_\mu,L_\nu]\big)=0,\label{iff}
\end{equation}

On the other hand, recently several authors have made an extremely useful contribution to the analytical calculations of QFIM.
In particular, Liu \textit{et al.} provided an analytical expression of the QFIM determined only by the support of the density matrix \cite{Liu2014a}.
Based on the spectral decomposition of $\rho(\bm{\theta})$
\begin{align}
\rho(\bm{\theta})=\sum_i^s\lambda_i(\bm{\theta})|\psi_i({\bm{\theta}})\rangle\langle\psi_i({\bm{\theta}})|,\label{decomposition}
\end{align}
with $s$ being the rank of $\rho(\bm{\theta})$ ($s\leq d$), the QFIM can be divided into two separate contributions
\begin{equation}
\mathcal{F}_{\mu\nu}=\mathcal{F}_C+\mathcal{F}_Q,
\end{equation}
where
\begin{align}
\mathcal{F}_C=&\sum_{i=1}^s\frac{\partial_\mu\lambda_i\partial_\nu\lambda_i}{\lambda_i},\nonumber\\
\mathcal{F}_Q=&\sum_{i=1}^s4\lambda_i\bm{\mbox{Re}}\Delta_{\mu\nu}^i-\sum_{i,j=1}^s\frac{8\lambda_i\lambda_j}{\lambda_i+\lambda_j}\bm{\mbox{Re}}\Theta_{\mu\nu}^{ij}. \label{formula}
\end{align}
with $\Delta_{\mu\nu}^i=\langle\partial_\mu\psi_i|\partial_\nu\psi_i\rangle$ and $\Theta_{\mu\nu}^{ij}=\langle\partial_\mu\psi_i|\psi_j\rangle\langle\psi_j|\partial_\nu\psi_i\rangle$.
Hence it can be seen clearly that $\mathcal{F}_C$ is attributed to the classical contribution if we treat the set of nonzero eigenvalues as a genuine  probability distribution;
while $\mathcal{F}_Q$ is the purely quantum contribution determined by both eigenvalues and eigenvectors. Furthermore, we notice that $\mathcal{F}(\bm{\theta})=[\mathcal{F}_{\mu\nu}]$
is a real symmetric matrix and its diagonal element coincides with the analytical formula of the single-parameter case as we expect \cite{Liu2013,Zhang2013,Liu2014b}.
Keeping these technical tools in mind, we are now in a position to present our main results.

\section{QFIM in quantum cloning machines}\label{calculations}
As described in the introduction, we mainly focus on the generalized $d$-dimensional equatorial states of the form
\begin{equation}
|\psi(\bm{\phi})\rangle=\frac{1}{\sqrt{d}}\sum_{j=0}^{d-1}e^{i\phi_j}|j\rangle,\label{equatorial}
\end{equation}
where $\bm{\phi}=\{\phi_0,\phi_2,\ldots,\phi_{d-1}\}$, $\phi_j\in[0,2\pi)$, $j=0,\ldots,d-1$, and $\{|j\rangle\}$ is a complete
orthonormal basis of the $d$-dimensional Hilbert space. The overall phase cannot be estimated, so we can assume $\phi_0=0$.
It is remarkable that this set of states (\ref{equatorial}) can be generated by $d-1$ independent phase shifts with respect to the reference state
$|\psi(\bm{\phi}=0)\rangle=(1/\sqrt{d})\sum_{j=0}^{d-1}|j\rangle$,
by virtue of the unitary transformation \cite{Macchiavello2003,Mista2005}
\begin{equation}
\mathcal{U}(\bm{\phi})=|0\rangle\langle 0|+\sum_{j=1}^{d-1} e^{i\phi_j}|j\rangle\langle j|,
\end{equation}

As a warm up, we first evaluate the QFIM of this initial states. By the definition, the SLDs of pure state $\rho(\bm{\theta})=|\psi_{\bm{\theta}}\rangle\langle\psi_{\bm{\theta}}|$
can be represented as
\begin{equation}
L_\mu=2\partial_\mu\rho(\bm{\theta})=2(|\partial_\mu\psi_{\bm{\theta}}\rangle\langle\psi_{\bm{\theta}}|+|\psi_{\bm{\theta}}\rangle\langle\partial_\mu\psi_{\bm{\theta}}|),\label{SLD}
\end{equation}
with $|\partial_\mu\psi_{\bm{\theta}}\rangle$ denoting the partial derivative of $|\psi_{\bm{\theta}}\rangle$ with respect to $\theta_\mu$.
Moreover, the QFIM can be rewritten as
\begin{equation}
\mathcal{F}_{\mu\nu}=4\bm{\mbox{Re}}\langle\psi_{\bm{\theta}}|L_\mu L_\nu|\psi_{\bm{\theta}}\rangle,\label{QFIM}
\end{equation}
Substituting Eq. (\ref{SLD}) into Eq. (\ref{QFIM}), one can obtains
\begin{equation}
\mathcal{F}_{\mu\nu}=4\bm{\mbox{Re}}\langle\partial_\mu\psi_{\bm{\theta}}|\Pi|\partial_\nu\psi_{\bm{\theta}}\rangle,
\end{equation}
where $\Pi=\bm{\mbox{I}}-|\psi_{\bm{\theta}}\rangle\langle\psi_{\bm{\theta}}|$ is the projection operator onto the
orthogonal complement of $\rho(\bm{\theta})$. With the notations as defined above, for the generalized equatorial states (\ref{equatorial}),
one gets
\begin{gather}
\Delta_{\mu\nu}=\langle\partial_\mu\psi(\bm{\phi})|\partial_\nu\psi(\bm{\phi})\rangle=\frac{1}{d}\delta_{\mu\nu},\nonumber\\
\Theta_{\mu\nu}=\langle\partial_\mu\psi(\bm{\phi})|\psi(\bm{\phi})\rangle\langle\psi(\bm{\phi})|\partial_\nu\psi(\bm{\phi})\rangle=\frac{1}{d^2},
\end{gather}
Therefore, the QFIM for states (\ref{equatorial}) can be expressed as
\begin{equation}
\mathcal{F}_{\mu\nu}=4\left(\Delta_{\mu\nu}-\Theta_{\mu\nu}\right)=4\left(\frac{1}{d}\delta_{\mu\nu}-\frac{1}{d^2}\right),
\end{equation}
Notice that $\Delta_{\mu\nu}$ and $\Theta_{\mu\nu}$ are all real-valued and $\bm{\mbox{Im}}[\langle\psi(\bm{\phi})|L_\mu L_\nu|\psi(\bm{\phi})\rangle]=0$
for all $\mu$ and $\nu$. Thus, the multi-parameter QCRB can be achieved in this case. Especially, the total variance of all the parameters follows
the inequality
\begin{equation}
(\Delta\bm{\phi})^2=\sum_{\mu=1}^{d-1}(\Delta\phi_\mu)^2=\textrm{Tr}[\textrm{Cov}(\bm{\phi})]\geq \textrm{Tr}[\mathcal{F}(\bm{\phi})^{-1}],
\end{equation}
We observe that in fact $\bm{\phi}$ is $d-1$ dimensional parameter vector and thus $\mathcal{F}(\bm{\phi})=[\mathcal{F}_{\mu\nu}]$ is a $d-1\otimes d-1$
matrix. According to the symmetry of $\mathcal{F}(\bm{\phi})$, the eigenvalues of $\mathcal{F}(\bm{\phi})^{-1}$ are $d^2/4$ and $d/4$, and the degrees
of degeneracy are $1$ and $d-2$ respectively. Therefore, the lower bound of the total variance is
\begin{equation}
(\Delta\bm{\phi})^2\geq\frac{d^2}{4}+\frac{d(d-2)}{4}=\frac{d(d-1)}{2}.
\end{equation}
Moreover, this error bound can indeed be achieved due to the saturation of the QCRB for the generalized equatorial states. Later, we are moving
on to the evaluation of the QFIMs of two essential types of quantum cloning machines.

\subsection{UQCM}
The UQCM was first proposed by Bu\u{z}ek and Hillery, in order to clone an arbitrary qubit to two approximate copies \cite{Buzek1996}.
The \textit{universality} indicates that the quality of the copies does not depend on the specific form of the input state. In other words,
all states should be copied equally well referring to a proper measure of the distance between the input and output states.
This cloning procedure was proved to be optimal, in the sense that the fidelity between the input qubit and output qubit is maximal \cite{Gisin1997,Bruss1998}.
Bu\u{z}ek and Hillery also extended the UQCM to the arbitrary-dimensional case, that is, $1\rightarrow2$ symmetric cloning of qudits \cite{Buzek1998}.

For a $d$-dimensional quantum system, the corresponding cloning mechanism can be specified as the following unitary transformation \cite{Buzek1998}
\begin{equation}
|i\rangle|0\rangle|X\rangle\Longrightarrow\alpha|i\rangle|i\rangle|X_i\rangle+\beta\sum_{i\neq j}^d(|i\rangle|j\rangle+|j\rangle|i\rangle)|X_j\rangle,
\end{equation}
where
\begin{equation}
\alpha=\frac{2}{\sqrt{2(d+1)}},\ \beta=\frac{1}{\sqrt{2(d+1)}},
\end{equation}
and $|i\rangle|0\rangle|X\rangle$ represent respectively the states of the original, the blank copy and the cloner qudit.
Here $\{|X_i\rangle\}$ denotes an orthonormal basis of the cloning machine Hilbert space.
It is worth noting that the UQCM can be completely characterized by a shrinking factor $\eta$ \cite{Bruss1998a} and it is
useful to express the output reduced state in the following form \cite{Buzek1998}
\begin{equation}
\rho^{\textrm{out}}=\eta\rho^{\textrm{in}}+\frac{1-\eta}{d}\bm{\mbox{I}},\label{shrinking}
\end{equation}
where $\rho^{\textrm{in}}=|\varphi\rangle\langle\varphi|$ describes the initial pure state to be cloned. It is easy to verify
that this scaling form indeed guarantees that the UQCM is input-state independent. Considering the equatorial states (\ref{equatorial})
as the input state, one of the two output qudits can be represented as
\begin{equation}
\rho^{\textrm{out}}(\bm{\phi})=\frac{d+2}{2(d+1)}|\psi(\bm{\phi})\rangle\langle\psi(\bm{\phi})|+\frac{1}{2(d+1)}\bm{\mbox{I}},\label{scaling}
\end{equation}

To apply the analytical formula presented in Eq. (\ref{formula}), our main task is to find the spectral decomposition of the mixed state (\ref{scaling})
(e.g., the diagonalization of $\rho^{\textrm{out}}(\bm{\phi})$). First, we observe that $|\psi(\bm{\phi})\rangle\langle\psi(\bm{\phi})|$ itself
is an eigenstate of $\rho^{\textrm{out}}(\bm{\phi})$, that is
\begin{equation}
\rho^{\textrm{out}}(\bm{\phi})|\psi\rangle\langle\psi|=\frac{d+3}{2(d+1)}|\psi\rangle\langle\psi|,
\end{equation}
Here and henceforth we omit the $\bm{\phi}$-dependence in $|\psi(\bm{\phi})\rangle$ for brevity. Therefore, the form (\ref{scaling}) can be recast as
\begin{equation}
\rho^{\textrm{out}}(\bm{\phi})=\frac{d+3}{2(d+1)}|\psi\rangle\langle\psi|+\frac{1}{2(d+1)}(\bm{\mbox{I}}-|\psi\rangle\langle\psi|),
\end{equation}
Now the problem is converted into the decomposition of the operator $\Pi=\bm{\mbox{I}}-|\psi\rangle\langle\psi|$ which is projected onto the
orthogonal complement of $|\psi\rangle\langle\psi|$. One possible set of orthonormal basis vectors of this $d-1$ dimensional Hilbert subspace
can be constructed as
\begin{equation}
|\psi_n\rangle=\sqrt{\frac{2n}{n+1}}\left(|\chi_n\rangle-\frac{1}{n}\sum_{j=1}^{n-1}e^{i\phi_{jn}}|\chi_j\rangle\right),\label{set2}
\end{equation}
where
\begin{equation}
|\chi_n\rangle=\frac{1}{\sqrt{2}}(-e^{-i\phi_{n0}},\ldots,\underbrace{1}_{\textrm{nth}},\ldots),
\end{equation}
Here we introduce the notation $\phi_{mn}=\phi_m-\phi_n$ and only the $0th$ and $nth$ ($1\leq n \leq d-1$) elements of $|\chi_n\rangle$ are nonzero (that is,
all $\ldots$ represent zeros). For more details, see the Appendix \ref{orthonormalization}.

From the above analysis, we finally obtain the spectral decomposition of $\rho^{\textrm{out}}(\bm{\phi})$
\begin{equation}
\rho^{\textrm{out}}(\bm{\phi})=\frac{d+3}{2(d+1)}|\psi_0\rangle\langle\psi_0|+\frac{1}{2(d+1)}\sum_{j=1}^{d-1}|\psi_j\rangle\langle\psi_j|,
\end{equation}
where we define $|\psi_0\rangle=|\psi(\bm{\phi})\rangle$ since $\{|\psi_n\rangle\}_{n=0}^{d-1}$ is exactly an orthonormal basis of the whole Hilbert space.
Combining the analytical formula (\ref{formula}) and this particular form of spectral decomposition, the diagonal elements of the QFIM are the same and can be evaluated as
(see the Appendix \ref{UQCM})
\begin{equation}
\mathcal{F}_{\mu\mu}^{\textrm{UQCM}}=\frac{2(d-1)(d+2)^2}{(d+1)(d+4)d^2},\label{diagonal}
\end{equation}
where $\mu=1,\ldots,d-1$. Correspondingly, the off-diagonal terms of the QFIM are also equal
\begin{equation}
\mathcal{F}_{\mu\nu}^{\textrm{UQCM}}=-\frac{2(d+2)^2}{(d+1)(d+4)d^2},\ (\mu\neq\nu)\label{off-diagonal}
\end{equation}

\begin{figure}[htbp]
\begin{center}
\includegraphics[width=.4\textwidth]{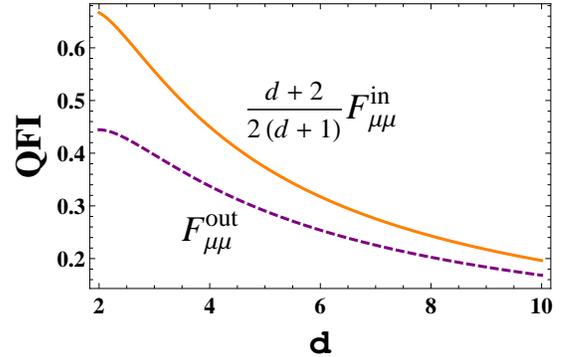} {}
\end{center}
\caption{(Color online) The confirmation of the inequality (\ref{inequality}). $\mathcal{F}_{\mu\mu}^{\textrm{in}}$ (orange solid line) and
$\mathcal{F}_{\mu\mu}^{\textrm{out}}$ (purple dashed line) represent the diagonal entries of the QFIM for the input state $|\psi(\bm{\phi})\rangle\langle\psi(\bm{\phi})|$
and output state $\rho^{\textrm{out}}(\bm{\phi})$, respectively.}
\label{f1}
\end{figure}

Before proceeding, some remarks need to be made. First, when $d=2$, Eq. (\ref{diagonal}) reduces to $\mathcal{F}_{11}=4/9$, which
recovers the qubit case presented in Ref. \cite{Song2013}. Secondly, for the initial pure state $|\psi(\bm{\phi})\rangle$, we notice that
the following relation holds
\begin{equation}
\mathcal{F}_{\mu\mu}=-(d-1)\mathcal{F}_{\mu\nu},\ (\mu\neq\nu) \label{relation}
\end{equation}
Intriguingly, this relation is still valid for the output mixed state $\rho^{\textrm{out}}(\bm{\phi})$ due to the scaling form (\ref{scaling}).
Finally, since a cloning scenario is a special kind of quantum channel (i.e., a trace-preserving completely positive map) \cite{DAriano2003},
QFI is non-increasing under the cloning transformation as a result of its monotonicity \cite{Fujiwara2001a}, that is
\begin{equation}
\mathcal{F}\left(\rho^{\textrm{out}}(\bm{\phi})\right)\leq\mathcal{F}\left(|\psi(\bm{\phi})\rangle\langle\psi(\bm{\phi})|\right),
\end{equation}
However, this inequality can be further strengthened combining the convexity of QFI and the scaling form of $\rho^{\textrm{out}}(\bm{\phi})$
\begin{equation}
\mathcal{F}\left(\rho^{\textrm{out}}(\bm{\phi})\right)\leq\frac{d+2}{2(d+1)}\mathcal{F}\left(|\psi(\bm{\phi})\rangle\langle\psi(\bm{\phi})|\right),
\end{equation}
Since a necessary condition for a real symmetric matrix to be positive is the positive definiteness of its diagonal entries, the following inequality
should be satisfied
\begin{equation}
\mathcal{F}_{\mu\mu}\left(\rho^{\textrm{out}}(\bm{\phi})\right)\leq\frac{d+2}{2(d+1)}\mathcal{F}_{\mu\mu}\left(|\psi(\bm{\phi})\rangle\langle\psi(\bm{\phi})|\right),\label{inequality}
\end{equation}
which is clearly confirmed by Fig. \ref{f1}.

\subsection{PQCM}
As described above, the UQCM is the optimal choice when the input state is completely unknown. However, in many realistic quantum information processing tasks,
we actually have a limited knowledge of the input state. By virtue of these partial information, a quantum cloning machine with better performance can be designed
for such a restricted class of input states. The first PQCM was proposed by D. Bru{\ss} \textit{et al.} for the equatorial qubit states of the form
$|\psi\rangle=(|0\rangle+e^{i\phi}|1\rangle)/\sqrt{2}$ \cite{Bruss1998}. Here \textit{phase-covariant} reveals that the quality of this cloning machine does not
rely on the specific values of phase parameter $\phi$. Then H. Fan \textit{et al.} presented explicitly the optimal $1\rightarrow M$ cloning transformation for
equatorial qubits \cite{Fan2001} and extended the PQCM to the $d$-dimensional quantum system \cite{Fan2003}.

Focusing on the generalized equatorial pure qudits (\ref{equatorial}), the optimal $1\rightarrow 2$ PQCM is characterized by the following unitary transformation \cite{Fan2003}
\begin{equation}
U|j\rangle|Q\rangle=\alpha|jj\rangle|R_j\rangle+\frac{\beta}{\sqrt{2(d-1)}}\sum_{l\neq j}^{d-1}(|jl\rangle+|lj\rangle)|R_l\rangle,
\end{equation}
where $|Q\rangle$ is a combination of the blank state and initial state of the cloning machine, $\{R_j\}$ is an orthonormal basis of the cloning machine and
\begin{align}
\alpha=\left(\frac{1}{2}-\frac{d-2}{2\sqrt{d^2+4d-4}}\right)^{1/2},\nonumber\\
\beta=\left(\frac{1}{2}+\frac{d-2}{2\sqrt{d^2+4d-4}}\right)^{1/2}.
\end{align}
By tracing over one qubit, we can obtain the reduced density matrix of a single output qudit
\begin{align}
\rho^{\textrm{out}}(\bm{\phi})=&\frac{1}{d}\sum_j|j\rangle\langle j|+\Bigg(\frac{\alpha\beta}{d}\sqrt{\frac{2}{d-1}}\nonumber\\
&+\frac{\beta^2(d-2)}{2d(d-1)}\Bigg)\sum_{j\neq k}e^{\phi_j-\phi_k}|j\rangle\langle k|,\label{reduced-qudit}
\end{align}
Remarkably, we notice that this output reduced state in Eq. (\ref{reduced-qudit}) can also be rewritten in the scaling form (\ref{shrinking})
with the shrinking factor
\begin{equation}
\eta^{PQCM}=\frac{1}{4(d-1)}\left(d-2+\sqrt{d^2+4d-4}\right),
\end{equation}
Since
\begin{equation}
\eta^{PQCM}>\eta^{UQCM}=\frac{d+2}{2(d+1)},
\end{equation}
the optimal fidelity of PQCM is larger than that of UQCM \cite{Fan2003}.

\begin{figure}[htbp]
\begin{center}
\includegraphics[width=.4\textwidth]{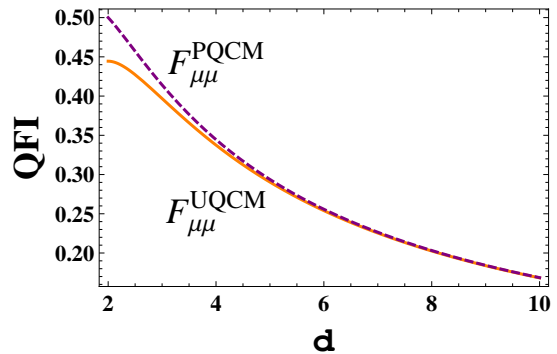} {}
\end{center}
\caption{(Color online) The comparison of the diagonal entries of QFIMs. $\mathcal{F}_{\mu\mu}^{\textrm{UQCM}}$ (orange solid line) and
$\mathcal{F}_{\mu\mu}^{\textrm{PQCM}}$ (purple dashed line) correspond to UQCM and PQCM, respectively.}
\label{f2}
\end{figure}

Following the same method as in the above section, we obtain the diagonal entries of the QFIM in this scenario
\begin{equation}
\mathcal{F}_{\mu\mu}^{\textrm{PQCM}}=\frac{2\left(d^2+d\gamma-2\gamma\right)}{d\big[d^2+d(\gamma+4)-2(\gamma+2)\big]},
\end{equation}
where $\gamma=\sqrt{d^2+4d-4}$. When $d=2$, $\mathcal{F}_{\mu\mu}^{\textrm{PQCM}}=1/2>4/9$.
Meanwhile, we observe that the relation (\ref{relation}) still holds in this circumstance.
Notably, it is easy to prove the inequality
\begin{equation}
\mathcal{F}_{\mu\mu}^{\textrm{PQCM}}\geq\mathcal{F}_{\mu\mu}^{\textrm{UQCM}},
\end{equation}
which means that the performance of PQCM is better than that of UQCM in terms of cloning QFI for each individual phase parameters.
However, when the dimensionality $d$ is large (e.g., $d\geq10$), it should be noted that the advantage of PQCM over UQCM almost disappears as shown in Fig. \ref{f2}.
This fact tells us that the PQCM is more significant for the qubit case. Furthermore, owing to the structure of QFIM and the relation (\ref{relation}), a stronger
(matrix) inequality holds (see the Appendix \ref{error})
\begin{equation}
\mathcal{F}^{\textrm{PQCM}}\geq\mathcal{F}^{\textrm{UQCM}}.
\end{equation}

\subsection{Generalization}
In fact, our method can be extended to any quantum cloning machines in which the output reduced state can be written in the form (\ref{shrinking}), that is
\begin{equation}
\rho^{\textrm{out}}(\bm{\phi})=\eta|\psi(\bm{\phi})\rangle\langle\psi(\bm{\phi})|+\frac{1-\eta}{d}\bm{\mbox{I}},
\end{equation}
where the shrinking factor $\eta$ does not depend on $|\psi(\bm{\phi})\rangle$.
The diagonal and off-diagonal elements of the QFIM for this general form of mixed qudit are given by
\begin{align}
\mathcal{F}_{\mu\mu}=&\frac{4(d-1)\eta^2}{d\big[2+(d-2)\eta\big]},\label{element1}\\
\mathcal{F}_{\mu\nu}=&-\frac{4\eta^2}{d\big[2+(d-2)\eta\big]},\ (\mu\neq\nu)\label{element2}
\end{align}
Therefore, we finally confirm that the relation $\mathcal{F}_{\mu\mu}=-(d-1)\mathcal{F}_{\mu\nu}$ is always valid due to both the structure of the initial state (\ref{equatorial})
and the scaling form of $\rho^{\textrm{out}}(\bm{\phi})$.

In addition, we find that $\mathcal{F}_{\mu\mu}$ is a monotonically increasing function of the shrinking factor $\eta$. Indeed, the first order derivative of $\mathcal{F}_{\mu\mu}$ is given by
\begin{equation}
\frac{\partial\mathcal{F}_{\mu\mu}}{\partial\eta}=\frac{4\eta(d-1)\big[4+(d-2)\eta\big]}{d\big[2+(d-2)\eta\big]^2}>0.
\end{equation}
This is to be expected since the larger $\eta$ is, the more information the reduced output state $\rho^{\textrm{out}}(\bm{\phi})$ contains about parameters.
Meanwhile, this finding also reconfirms the previous result that $\mathcal{F}_{\mu\mu}^{\textrm{PQCM}}\geq\mathcal{F}_{\mu\mu}^{\textrm{UQCM}}$ since $\eta^{PQCM}>\eta^{UQCM}$.

Moreover, it should be emphasised that the structure of the QFIM is heavily dependent on the form of the input state $|\psi(\bm{\phi})\rangle$.
Here we are focusing on the generalized equatorial states and this the reason why the diagonal (or off-diagonal) entries are all equal.
When the parameters are encoded in the initial state in a more complex way, a more technical treatment will be involved but the critical
point is still to diagonalize the reduced state $\rho^{\textrm{out}}(\bm{\phi})$.

\section{Attainability of QCRB }\label{attainability}
For the ideal pure state (\ref{equatorial}), the multi-parameter QCRB can be saturated, that is, the optimal measurements performed to attain the
quantum limits for every individual parameters commute with each other. To identify whether the QCRB can be achieved for the output reduced state
$\rho^{\textrm{out}}(\bm{\phi})$, we should check the condition (\ref{iff}) for every pair of SLDs. However, it could be a very difficult task to
apply this criteria directly since the explicit expression of SLD is usually hard to obtain. Similar to the formula (\ref{formula}), here we present
an analytical expression of this criteria exploiting the diagonalization of $\rho^{\textrm{out}}(\bm{\phi})$ (see the Appendix \ref{error})
\begin{align}
\textrm{Tr}\left(\rho(\bm{\phi})\frac{[L_\mu,L_\nu]}{2}\right)=&i\Bigg(\sum_{k=1}^s4\lambda_{k}\bm{\mbox{Im}}\Delta_{\mu\nu}^k \nonumber\\
&-\sum_{k,l=1}^s\frac{8\lambda_k\lambda_l(\lambda_k-\lambda_l)}{(\lambda_k+\lambda_l)^2}\bm{\mbox{Im}}\Theta_{\mu\nu}^{kl}\Bigg),
\end{align}
In fact, $\Delta_{\mu\nu}^k$ and $\Theta_{\mu\nu}^{kl}$ are all real-valued based on our construction. Therefore, the multi-parameter QCRB
is attainable in our study.

\begin{figure}[htbp]
\begin{center}
\includegraphics[width=.4\textwidth]{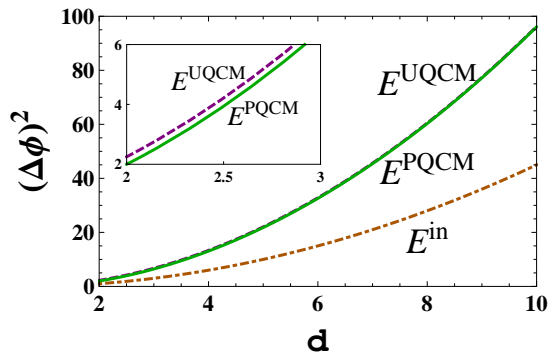} {}
\end{center}
\caption{(Color online) The total variances (errors) for multiple phase estimation. $E^{\textrm{in}}$ (orange dot-dashed line), $E^{\textrm{UQCM}}$ (green solid line)
and $E^{\textrm{PQCM}}$ (purple dashed line) represent the total \textit{errors} for quantum simultaneous estimation of all the phases using the initial pure state, the output reduced state
of UQCM and PQCM, respectively. The inset picture clearly shows that $E^{\textrm{UQCM}}>E^{\textrm{PQCM}}$.}
\label{f3}
\end{figure}

On the other hand,  for the output reduced state $\rho^{\textrm{out}}(\bm{\phi})$, the total variance (error) of all the phases $\{\phi_\mu\}_{\mu=1}^{d-1}$
is lower bounded by
\begin{equation}
(\Delta\bm{\phi})^2=\textrm{Tr}[\mbox{Cov}(\bm{\phi})]\geq \textrm{Tr}[\mathcal{F}(\bm{\phi})^{-1}],
\end{equation}
Because of the saturation of the matrix QCRB , this lower bound can also be achieved. Form Eqs. (\ref{element1}) and (\ref{element2}),
the analytical expression of this lower bound can be obtained (see the Appendix \ref{error})
\begin{equation}
(\Delta\bm{\phi})^2_{\textrm{min}}=\textrm{Tr}[\mathcal{F}(\bm{\phi})^{-1}]=\frac{(d-1)\big[2+(d-2)\eta\big]}{2\eta^2}.
\end{equation}
Remember that $\mathcal{F}(\bm{\phi})=[\mathcal{F}_{\mu\nu}]$ is a $d-1\otimes d-1$ matrix. As shown in Fig. \ref{f3}, for the purpose of simultaneously estimating
all the phases, the PQCM has an advantage over the UQCM, since the total error $(\Delta\bm{\phi})^2_{\textrm{min}}$ is a monotonically decreasing function of the shrinking factor $\eta$.
In particular, when $\eta=1$, we recover the result for the initial pure state. Nevertheless, it is also evident that this advantage is not very significant, as seen from from Fig. \ref{f3}.
In fact, to see this, we notice that when $d\rightarrow\infty$ both of the output reduced states of UQCM and PQCM asymptotically approach an \textit{identical} final state since
\begin{equation}
\lim_{d\rightarrow\infty}\eta^{UQCM}=\lim_{d\rightarrow\infty}\eta^{PQCM}=\frac{1}{2}.
\end{equation}

\section{DISCUSSION AND CONCLUSION}\label{conclusion}
In contrast to the single parameter issue, recently increasing attention has been paid to the multiple parameter estimation problem, especially
from quantum information perspective. On one hand, in many practical scenarios, more than one parameters are naturally involved and the simultaneous
estimation of these parameters is of significant interest to the research community on both theoretical and experimental grounds.
On the other hand, due to the quantum nature, quantum estimation of multiple parameters is fundamentally distinct from the single parameter case,
since the SLDs corresponding to different parameter do not commute with each other in general (which means the optimal measurements for each individual
parameters are incompatible). In addition to these basic considerations, we realize that quantum cloning of high-dimensional systems can be regarded as
a multi-parameter estimation problem and it provides an excellent platform for investigating the quantum feature of this scenario.

In this study, we concentrate on the generalized $d$-dimensional equatorial qudit as the input state, not only due to its symmetry but also for
its importance in quantum information processing tasks. Within the framework of quantum cloning machines, we present the analytical expressions
of the QFIMs for UQCM and PQCM, and prove that PQCM indeed performs better than UQCM in terms of QFI-cloning. It is also worth emphasizing that
our method can be directly extended to any cloning machines where the output reduced state can be written as the scaling form (\ref{shrinking}).
When dealing with the attainability of QCRB, we introduce a new matrix $\mathcal{L}(\bm{\theta})=[\mathcal{L}_{\mu\nu}]$ (see Appendix \ref{error}),
which is dual to $\mathcal{F}(\bm{\theta})$ and directly determines whether the QCRB can be achieved. We provide an analytical formula for elements
of $\mathcal{L}(\bm{\theta})$ and show that the ultimate quantum limits can be attained in our study.

Based on these findings, a wider variety of problems deserve our attention: (i) the multi-parameter estimation strategies need to be investigated
under the background of other quantum cloning scenarios, such as the state-dependent cloning \cite{Bruss1998} and probabilistic quantum cloning machines \cite{Duan1998a,Duan1998b}.
Especially for the latter, a post-selection of the measurement results is involved and the role played by post-selection in quantum metrology
has attracted a lot of attention recently \cite{Tanaka2013,Knee2013,Knee2014,Ferrie2014,Combes2014}. Mixed-state cloning or broadcasting would
be also very interesting in terms of QFI \cite{Lu2013}. (ii) As pointed out by Ref. \cite{Song2013}, we could directly take the QFI as the figure of
merit to find out the \textit{optimal} cloning machines in this context, particularly when it is unnecessary to acquire the full information of
quantum states.

\begin{acknowledgments}
Y. Yao and L. Ge contributed equally to this work.
The authors are supported by the National Natural Science Foundation of China (Grants No. 11121403, No. 10935010, No. 11074261, and No. 11247006),
the National 973 program (Grants No. 2012CB922104 and No. 2014CB921403),
and the China Postdoctoral Science Foundation (Grant No. 2014M550598).
\end{acknowledgments}
\appendix

\section{The choice of decomposition and Gram-Schmidt orthonormalization}\label{orthonormalization}
To be clear, our main concern is to construct a complete set of orthonormal basis vectors of the orthogonal complement to $|\psi(\bm{\phi})\rangle\langle\psi(\bm{\phi})|$.
The first step is to find $d-1$ vectors which span this $d-1$ dimensional Hilbert subspace, although they may not be orthogonal to each other. The general form of pure qudit can be expressed as
\begin{equation}
|\chi\rangle=\sum_{j=0}^{d-1}\alpha_j|j\rangle,
\end{equation}
where $\alpha_j$ are complex coefficients and $\sum_{j=0}^{d-1}|\alpha_j|^2=1$. Since these vectors are orthogonal to $|\psi(\bm{\phi})\rangle\langle\psi(\bm{\phi})|$,
they should satisfy the following condition
\begin{equation}
\langle\chi|\psi(\bm{\phi})\rangle=\frac{1}{\sqrt{d}}\sum_{j=0}^{d-1}\alpha_j^\ast e^{i\phi_j}=0,
\end{equation}

Intuitively, the simplest form of $|\chi\rangle$ is what we presents in the main text, that is
\begin{equation}
|\chi_n\rangle=\frac{1}{\sqrt{2}}(-e^{-i\phi_{n0}},\ldots,\underbrace{1}_{\textrm{nth}},\ldots),\label{set1}
\end{equation}
where $1\leq n \leq d-1$ and all $\ldots$ represent zeros. In fact, a more general form can be given as
\begin{equation}
|\chi'_n\rangle=\frac{1}{\sqrt{2}}(\ldots,\underbrace{-e^{-i\phi_{nm}}}_{\textrm{mth}},\ldots,\underbrace{1}_{\textrm{nth}},\ldots),
\end{equation}
where $\phi_{nm}=\phi_{n}-\phi_{m}$ and $m$ is free to choose with $m<n$. Moreover, we should keep in mind that the rule of inner products of vectors (\ref{set1}) is
\begin{align}
\left\{\begin{array}{cc}
\langle\chi_m|\chi_n\rangle=\frac{1}{2}e^{i\phi_{mn}}, & \mbox{ if } \, m\neq n\\
\langle\chi_m|\chi_n\rangle=1, & \mbox{ if } \, m=n
\end{array}\right.\label{rule}
\end{align}

However, Eqs. (\ref{rule}) implies that $|\chi_n\rangle$ is not orthogonal to each other.
To get an orthonormal basis of this Hilbert subspace, we need to make use of the procedure of Gram-Schmidt orthonormalization.
The Gram-Schmidt process generates an orthogonal set of vectors $\Omega'=\{|\omega_1\rangle,\ldots,|\omega_{d}\rangle\}$ from
a finite linearly independent set $\Omega=\{|\upsilon_1\rangle,\ldots,|\upsilon_{d}\rangle\}$ which span the the same $d$-dimensional subspace.
Defining $|\xi_1\rangle=|\upsilon_1\rangle/\||\upsilon_1\rangle\|$, the Gram-Schmidt process works inductively as follows
\begin{equation}
|\omega_k\rangle=|\upsilon_k\rangle-\sum_{i=1}^{k-1}\langle\upsilon_k|\xi_i\rangle|\xi_i\rangle,\ |\xi_k\rangle=\frac{|\omega_k\rangle}{\||\omega_k\rangle\|}
\end{equation}
where $2\leq k\leq d$ and $\{|\xi_1\rangle,\ldots,|\xi_{d}\rangle\}$ is the required set of normalized orthogonal vectors.

Moving on to our case and utilizing the rules in Eqs. (\ref{rule}), the Gram-Schmidt process produces a sequence of \textit{unnormalized} vectors
\begin{equation}
|\widetilde{\psi}_n\rangle=|\chi_n\rangle-\frac{1}{n}\sum_{j=1}^{n-1}e^{i\phi_{jn}}|\chi_j\rangle,
\end{equation}
with $1\leq n\leq d-1$. Through direct calculation, we notice that
\begin{equation}
\langle\widetilde{\psi}_n|\widetilde{\psi}_n\rangle=\frac{n+1}{2n},
\end{equation}
Therefore, after the normalization, the desired set of vectors is just as the states (\ref{set2}) given in the main text.
As expected, one can easily check that $\langle\psi_m|\psi_n\rangle=\delta_{mn}$. Moreover, it is worth emphasizing
that this choice of decomposition does not lose any generality, since distinct sets of orthonormal basis are related by
unitary transformations and QFI is invariant under unitary transformations.

\section{QFIM for UQCM}\label{UQCM}
First, we observe that there is no classical contribution (see the formula (\ref{formula})), since the eigenvalues
contain no information about $\bm{\phi}$
\begin{equation}
\lambda_0=\frac{d+3}{2(d+1)},\ \lambda_n=\frac{1}{2(d+1)},
\end{equation}
with $1\leq n\leq d-1$. Before evaluating the quantum part, there are two points which need to be clarified:
(i) Due to the symmetry of $|\psi(\bm{\phi})\rangle$ and the scaling form of $\rho^{\textrm{out}}(\bm{\phi})$,
all $\{\phi_\mu\}_{\mu=1}^{d-1}$ are encoded in $\rho^{\textrm{out}}(\bm{\phi})$ on an equal footing. More precisely,
the diagonal (or off-diagonal) elements of the QFIM will show a \textit{similar dependence} on the set of parameters. For instance,
if we find $\mathcal{F}_{11}$ is independent of all the parameters, then all $\mathcal{F}_{\mu\mu}$ will be all equal and have
no dependence on any $\phi_\mu$ (later we will prove this is indeed the case); (ii) The quantum contribution is composed of two isolated terms
and these two summations can be calculated separately. The key issue is to determine $\Delta_{\mu\nu}$ and $\Theta_{\mu\nu}$ for certain parameters.

In the following, we try to evaluate $\mathcal{F}_{11}$, that is, $\mu=\nu=1$. Based on the orthonormal basis $\{|\psi_n\rangle\}_{n=0}^{d-1}$ and
defining $\Delta_{\mu\nu}^n=\langle\partial_\mu\psi_n|\partial_\nu\psi_n\rangle$, we have
\begin{align}
\Delta_{\mu\nu}^n=\left\{\begin{array}{cc}
1/d, & \mbox{ if } \, n=0\\
1/n(n+1), & \mbox{ if } \, 1\leq n\leq d-1
\end{array}\right.
\end{align}
Thus the first summation is
\begin{equation}
\sum_{n=0}^{d-1}4\lambda_n\bm{\mbox{Re}}\Delta_{\mu\nu}^n=\frac{4}{d},\label{term1}
\end{equation}
where we make use of the identity
\begin{equation}
\sum_{n=1}^{d-1}\frac{1}{n(n+1)}=1-\frac{1}{d},
\end{equation}
On the other hand, it is much more complicated to calculate $\Theta_{\mu\nu}^{nm}=\langle\partial_\mu\psi_n|\psi_m\rangle\langle\psi_m|\partial_\nu\psi_n\rangle$.
Here we only present the results
\begin{align}
\Theta_{\mu\nu}^{nm}=\left\{\begin{array}{cccc}
\frac{1}{d^2}, & \mbox{ if } \, n=m=0\\
\frac{1}{dm(m+1)}, & \mbox{ if } \, n=0,m\geq1\\
\frac{1}{dn(n+1)}, & \mbox{ if } \, n\geq1,m=0\\
\frac{1}{nm(n+1)(m+1)}, & \mbox{ if } \, n\geq1,m\geq1
\end{array}\right.
\end{align}
Therefore, we obtain the second term
\begin{equation}
\sum_{n,m=0}^{d-1}\frac{8\lambda_n\lambda_m}{\lambda_n+\lambda_m}\bm{\mbox{Re}}\Theta_{\mu\nu}^{nm}=\frac{2(d^3+7d^2+8d+4)}{(d+1)(d+4)d^2},\label{term2}
\end{equation}
Subtracting Eq. (\ref{term2}) from Eq. (\ref{term1}), we obtain $\mathcal{F}_{11}$ and it is indeed independent of any parameter. Therefore,
all the diagonal elements are equal to $\mathcal{F}_{11}$. Following a similar procedure as above, we can also evaluate the off-diagonal
terms of the QFIM (see Eq. (\ref{off-diagonal})). The calculations are tedious but straightforward, and so the details are not
presented here for the sake of simplicity.

\section{Attainability}\label{error}
Following the notations in Ref. \cite{Liu2014a}, the elements of QFIM are defined by
\begin{equation}
\mathcal{F}_{\mu\nu}=\frac{1}{2}\textrm{Tr}\big[\rho(\bm{\theta})\{L_\mu,L_\nu\}\big],
\end{equation}
Correspondingly, here we introduce another matrix $\mathcal{L}(\bm{\theta})=[\mathcal{L}_{\mu\nu}]$, whose elements read
\begin{equation}
\mathcal{L}_{\mu\nu}=\frac{1}{2}\textrm{Tr}\big[\rho(\bm{\theta})[L_\mu,L_\nu]\big],
\end{equation}
In fact, one can find that
\begin{align}
\mathcal{F}_{\mu\nu}=\bm{\mbox{Re}}\textrm{Tr}\big[\rho(\bm{\theta}) L_\mu L_\nu\big],\\
\mathcal{L}_{\mu\nu}=i\bm{\mbox{Im}}\textrm{Tr}\big[\rho(\bm{\theta}) L_\mu L_\nu\big],
\end{align}

Moreover, based on the spectral decomposition (\ref{decomposition}), the elements of $\mathcal{L}(\bm{\theta})$
can be represented as
\begin{equation}
\mathcal{L}_{\mu\nu}=\frac{1}{2}\sum_{i=1}^s\sum_{j=1}^d\lambda_i\big([L_\mu]_{ij}[L_\nu]_{ji}-[L_\nu]_{ij}[L_\mu]_{ji}\big),
\end{equation}
where we define $[L_\mu]_{ij}=\langle\psi_i|L_\mu|\psi_j\rangle$ and note that $[L_\mu]_{ij}=[L_\mu]_{ji}^\ast$.
Using results from Ref. \cite{Liu2014a}, one can find that
\begin{align}
\sum_{i=1}^s\sum_{j=1}^d\lambda_i[L_\mu]_{ij}[L_\nu]_{ji}=&\sum_{i=1}^s\frac{\partial_\mu\lambda_i\partial_\nu\lambda_i}{\lambda_i}+\sum_{i=1}^s4\lambda_i\Delta_{\mu\nu}^i\nonumber\\
&-\sum_{i,j=1}^s\frac{16\lambda_i^2\lambda_j}{(\lambda_i+\lambda_j)^2}\Theta_{\mu\nu}^{ij},
\end{align}
Therefore, we can ontain
\begin{align}
\mathcal{L}_{\mu\nu}=i\Bigg(\sum_{k=1}^s4\lambda_k\bm{\mbox{Im}}\Delta_{\mu\nu}^k-\sum_{k,l=1}^s\frac{16\lambda_k^2\lambda_l}{(\lambda_k+\lambda_l)^2}\bm{\mbox{Im}}\Theta_{\mu\nu}^{kl}\Bigg),
\end{align}
Here we should note the fact that
\begin{align}
\bm{\mbox{Re}}\Theta_{\mu\nu}^{kl}=&\bm{\mbox{Re}}\Theta_{\mu\nu}^{lk},\\
\bm{\mbox{Im}}\Theta_{\mu\nu}^{kl}=&-\bm{\mbox{Im}}\Theta_{\mu\nu}^{lk},
\end{align}
In fact, for an antisymmetric matrix $\mathcal{A}_{ij}=-\mathcal{A}_{ji}$, we have the relation
\begin{align}
\sum_{ij}\lambda_i\mathcal{A}_{ij}&=\frac{1}{2}\sum_{ij}(\lambda_i\mathcal{A}_{ij}+\lambda_j\mathcal{A}_{ji})\nonumber\\
&=\frac{1}{2}\sum_{ij}(\lambda_i-\lambda_j)\mathcal{A}_{ij}
\end{align}
Then we obtain the final expression in the main text. Remarkably, in contrast to the expression of $\mathcal{F}_{\mu\nu}$,
there is no classical contribution and this fact implies that whether $\mathcal{L}_{\mu\nu}$ ($\mu\neq\nu$) are equal to zero or not depends
on purely quantum effect.

On the other hand, the structure of QFIM is of the form
\begin{equation}
\mathcal{F}=
\left(\begin{array}{cccc}
\mathcal{F}_{\mu\mu} & \mathcal{F}_{\mu\nu} & \cdots & \mathcal{F}_{\mu\nu} \\
\mathcal{F}_{\mu\nu} & \mathcal{F}_{\mu\mu} & \cdots & \mathcal{F}_{\mu\nu} \\
\vdots & \vdots & \ddots & \vdots \\
\mathcal{F}_{\mu\nu} & \mathcal{F}_{\mu\nu} & \cdots & \mathcal{F}_{\mu\mu}
\end{array}\right),
\end{equation}
Thus the eigenvalues of $\mathcal{F}$ are given by
\begin{gather}
\lambda_1=\mathcal{F}_{\mu\mu}+(d-2)\mathcal{F}_{\mu\nu},\nonumber\\
\lambda_2=\cdots=\lambda_{d-1}=\mathcal{F}_{\mu\mu}-\mathcal{F}_{\mu\nu},
\end{gather}
From this result, one can easily obtain the lower bound of the total variance
\begin{align}
(\Delta\bm{\phi})^2_{\textrm{min}}=&\frac{1}{\mathcal{F}_{\mu\mu}+(d-2)\mathcal{F}_{\mu\nu}}+\frac{d-2}{\mathcal{F}_{\mu\mu}-\mathcal{F}_{\mu\nu}}\nonumber\\
=&-\frac{2(d-1)}{d\mathcal{F}_{\mu\nu}}\nonumber\\
=&\frac{(d-1)\big[2+(d-2)\eta\big]}{2\eta^2}
\end{align}
where the relation $\mathcal{F}_{\mu\mu}=-(d-1)\mathcal{F}_{\mu\nu}$ has been used.


\end{document}